\documentclass[12pt,graphicx]{article}
\usepackage{amssymb}

\usepackage{amsmath}
\usepackage[dvips]{graphics}
\usepackage[dvips]{graphicx}

\newcommand{\benumerate}{\begin{enumerate}}
\newcommand{\eenumerate}{\end{enumerate}}

\newcommand{\bitemize}{\begin{itemize}}

\newcommand{\eitemize}{\end{itemize}}
\newcommand{\der}[2]{\frac{\partial #1}{\partial #2}}

\newcommand{\paper}[6]{#1, #2, {\em #3}, {\bf #4}, #5,  (#6)}

\newcommand{\book}[4]{#1, {\em #2}, {#3}, (#4)}

\begin{document}

\date{}

\title{Light propagation in a Cole-Cole nonlinear medium via
  Burgers-Hopf equation.}

\author{Boris Konopelchenko\footnote{Supported in part by the COFIN
PRIN ``SINTESI'' 2002.}~ and
Antonio Moro$^{\ast}$ \\
{\em Dipartimento di Fisica dell'Universit\`{a} di Lecce} \\
{\em and INFN, Sezione di Lecce, I-73100 Lecce, Italy} \\
\footnotesize{E-mail: konopel@le.infn.it, antonio.moro@le.infn.it}}

\maketitle

\begin{abstract}
Recently, a new model of propagation of the light through
the so-called weakly three-dimensional Cole-Cole nonlinear medium with
short-range nonlocality has
been proposed. In particular, it has been shown that in the
geometrical optics limit, the model is integrable and it is governed
by the dispersionless Veselov-Novikov (dVN) equation.

Burgers-Hopf equation can be obtained as 1+1-dimensional reduction of
dVN equation. We discuss its properties in the specific context of nonlinear
geometrical optics. An illustrative explicit example is considered.

\end{abstract}

Many recent studies concerning the dispersionless integrable
systems have shown their relevance in a broad variety of fields in
physics such as Laplacian growth, topological field theory,
nonlinear optics~\cite{Interface_tau}-\cite{Moro2}, as well as
their deep connection with various fields in mathematics, in
particular, with the theory of asymptotic expansions, conformal and
quasiconformal mappings and the study of integrable deformations
of complex algebraic curves~\cite{LaxLev}-\cite{Konalgebraic}.

In the present paper we are interested in the application of the
theory of dispersionless integrable systems in nonlinear geometrical
optics. In particular, it has been shown in~\cite{Moro1} that
the Maxwell equations describing the propagation of the light through
the Cole-Cole weakly three-dimensional nonlinear medium admit an
integrable geometrical optics limit.

In this limit the system is governed by the dispersionless
Veselov-Novikov (dVN) hierarchy, which is amenable by the
quasiclassical $\bar{\partial}$-dressing method~\cite{Moro2}.

A reduction method based on the symmetry
constraints~\cite{Bogdanov1} appears to be an efficient approach
to calculate solutions of the dVN equation~\cite{Moro3}.

In what follows we will focus on a 1+1-dimensional reduction when
the refractive index does not depend on one coordinate. In this
case dVN hierarchy is reduced to the so-called Burgers-Hopf (BH)
hierarchy. For sake of simplicity we will focus on the first
equations of the hierarchy, the so-called dVN and Burgers-Hopf
equations, respectively.

In this regard, it is quite interesting to interpret some of the properties
of BH equation within the specific context of nonlinear geometrical
optics.
One of them concerns the existence of breaking wave solutions. They could be
useful to model dielectrics which present a kind of ``impurities'' inducing
abrupt variations of the refractive index. Indeed, it can be seen that, at
so-called breaking points, the curvature of the light rays blows up
just like it happens at interface between different media.

Let us start with the Maxwell equations in a dielectric medium
\begin{gather}
\label{maxwell}
\begin{aligned}
&\nabla \wedge {\bf H} - \der{{\bf D}}{t}=0 \quad{}\nabla \cdot {\bf D} = 0 \\
&\nabla \wedge {\bf E} + \der{{\bf B}}{t}=0 \quad{}\nabla \cdot
{\bf B} = 0
\end{aligned}
\end{gather}
along with the {\em constitutive equations}
\begin{equation}
\label{materials} {\bf D} = \varepsilon {\bf E},
\quad{}{\bf
B} = \mu {\bf H}.
\end{equation}
Looking for monochromatic solutions
\begin{gather}
\label{full_fields}
\begin{aligned}
 &{\bf E}\left(x,y,z,t\right) = {\bf
E_{0}}\left(x,y,z\right) e^{-i\omega t} \\
&{\bf H}\left(x,y,z,t\right) = {\bf
H_{0}}\left(x,y,z\right)e^{-i\omega t},
\end{aligned}
\end{gather}
one gets from~(\ref{maxwell}-\ref{materials}) the following well-known second order equations
\begin{gather}
\label{second}
\begin{aligned}
\nabla^{2}{\bf E_{0}} + \omega^{2} \mu
\varepsilon {\bf E_{0}} + \left (\nabla \log~\mu \right) \wedge
\left(\nabla \wedge {\bf E_{0}} \right ) +
\nabla \left ({\bf E_{0}} \cdot \nabla \log~\varepsilon \right) &= 0 \\
\nabla^{2}{\bf H_{0}} + \omega^{2} \mu \varepsilon {\bf H_{0}} +
\left (\nabla \log~\varepsilon \right) \wedge \left(\nabla \wedge
{\bf H_{0}} \right ) + \nabla \left ({\bf H_{0}} \cdot \nabla
\log~\mu \right) &= 0.
\end{aligned}
\end{gather}
Medium under study is characterized by the following set of
properties~\cite{Moro1,Moro2}:
\begin{enumerate}
\item $\varepsilon$ and $\mu$ obeys the {\em Cole-Cole dispersion law}
\begin{gather}
\label{ColeCole}
\begin{aligned}
\varepsilon = \varepsilon_{0} +
\frac{\tilde{\varepsilon}}{1+\left (i
  \omega \tau_{0} \right)^{2\nu}}, \\
\mu = \mu_{0} +
\frac{\tilde{\mu}}{1+\left (i
  \omega \tau_{0} \right)^{2\nu}},
\end{aligned}
\quad{}\quad{}0< \nu < \frac{1}{2}
\end{gather}
We highlight that the range of values of the exponent $\nu$ plays a crucial
role in the construction of an integrable high frequency limit.
\item $\varepsilon_{0}$ and $\mu_{0}$ depend only on the
  coordinates $x$, $y$ and $z$, while $\tilde{\varepsilon}$ and
  $\tilde{\mu}$ are assumed to be depending on the coordinates, the
  fields and the spatial derivatives of the fields. The latter can be
  explained in terms of a
mechanism of short-range non-locality by means of an integral constitutive
relation among the electric field $\bf{E}$ and the displacement vector
  $\bf{D}$~\cite{Moro2}.

\item All quantities, in high frequency limit show slow dependence on
  the variable $z$ formally given by 
\[
\der{}{z} = \omega^{-\nu} \der{}{\xi}.
\]
where $\xi$ is a ``slow'' variable defined by 
$z = \omega^{\nu} \xi$.

\noindent Moreover, any of such a quantity $f(x,y,z)$ can be expanded in
asymptotic series on the parameter $\omega^{\nu}$
\[
f(x,y,z) = f(x,y,\xi) + \omega^{-\nu} f_{1}(x,y,\xi) + \omega^{-2 \nu}
f_{2}(x,y,\xi) + \dots .
\]
\end{enumerate}

\noindent Representing
\begin{equation}
\label{fields}
{\bf E}_{0} = {\bf \tilde{E}}_{0} e^{i \omega S}, \quad{}{\bf H}_{0} =
{\bf \tilde{H}}_{0} e^{i \omega S},
\end{equation}
and using the previous assumptions in one of equations~(\ref{second}), in
high frequency limit ($\omega \to \infty$) one gets in the orders $\omega^{2}$ and
$\omega^{2 - 2 \nu}$ the main non-trivial
contributions~\cite{Moro1,Moro2}
\begin{align}
\label{dVN_phase1}
&S_{x}^{2} + S_{y}^{2} = 4 u, \\
\label{dVN_phase2}
&S_{\xi} = \varphi \left (x,y,\xi,S_{x},S_{y} \right ).
\end{align}
Equation~(\ref{dVN_phase1}) is the standard eikonal equation
in two dimensions, $\varphi$ is a certain function and $4 u = \varepsilon_{0} \mu_{0}$.
As discussed in the paper~\cite{Moro2} the equations~(\ref{dVN_phase1})
and~(\ref{dVN_phase2}) constitute an overdetermined system for
the phase $S$. The compatibility condition imposes specific
restrictions on the possible forms of the function $\varphi$ and the
admissible refractive indices $u$. If $\varphi$ is a polynomial
differential of $S$,
the first non-trivial case is given by the third degree polynomial,
i.e.
\begin{equation}
\varphi = \frac{1}{4} S_{x}^{3} - \frac{3}{4} S_{x} S_{y}^{2} + V_{1}
S_{x} + V_{2} S_{y}.
\end{equation}
The compatibility condition gives
\begin{gather}
\label{dVN}
\begin{aligned}
&u_{\xi} = \left(V_{1} u \right)_{x} + \left(V_{2} u \right)_{y} \\
&V_{1x} - V_{2y} = - 3 u_{x} \\
&V_{2x} + V_{1y} = 3 u_{y},
\end{aligned}
\end{gather}
which is the dVN equation. In the paper~\cite{Moro3} it has been
shown how symmetry constraints allow us to construct its 1+1-dimensional
reductions of hydrodynamic type. A number of explicit solutions
for the complex dVN equation have been discussed in~\cite{Moro2}.

Particular situation in which refractive index depend only on one
coordinate on the plane $x$-$y$ is of physical interest too. So,
let us assume that
\begin{equation}
\label{Dy}
u_{y} = 0.
\end{equation}
As a consequence of the eikonal equation~(\ref{dVN_phase1}) the phase
function $S$ must satisfy the condition $S_{y} = c$, with $c = \text{const}$,
then
\begin{equation}
S = c y + \tilde{S}(x,\xi)
\end{equation}
where $\tilde{S}$ does not depend on $y$. Eikonal
equation~(\ref{dVN_phase1}) becomes
\begin{equation}
\tilde{S}_{x}^{2} = 4 u - c^{2},
\end{equation}
while, choosing $V_{2} = 0$, one gets
\begin{equation*}
V_{1} = - 3 u,
\end{equation*}
and the dVN equation~(\ref{dVN}) is reduced to
the Burgers-Hopf (BH)
equation
\begin{equation}
\label{Burgers} u_{\xi} + 6 u u_{x} = 0.
\end{equation}
Solutions of BH equation can be expressed in the following hodograph
form~\cite{whithambook} 
\begin{equation}
\label{hodograph} x - 6 u \xi + \psi (u) = 0,
\end{equation}
where $\psi$ is an arbitrary function of its argument. Moreover,
once calculated $u(x,\xi)$ by the~(\ref{hodograph}) the phase
function is given explicitly
\begin{equation}
\label{phase} S = c y \pm \int \sqrt{4 u  - c^{2}} dx.
\end{equation}
Shock structure of the Burgers-Hopf equation is connected to the
existence of a value $\xi^{\ast}$ ({\em breaking point}), for which
\begin{equation}
u_{x} \to  \infty.
\end{equation}
\begin{figure}
\begin{center}
\includegraphics[width=12cm]{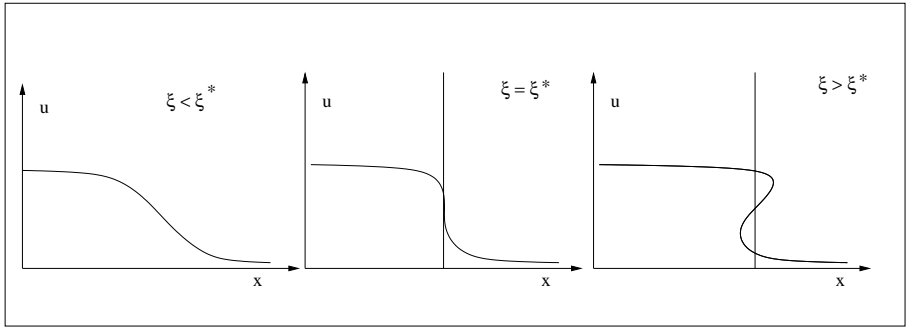}
\caption{\footnotesize{Typical breaking solution of Burgers-Hopf
    equation. Beyond breaking point ($\xi > \xi^{*}$) function $u$
    becomes multi-valued.}}
\label{break}
\end{center}
\end{figure}
In the theory of partial differential equations these solutions
are called {\em breaking waves}.
 It is a textbook exercise to calculate the breaking
 point~\cite{whithambook}. For instance, choosing the initial datum
 $u(x_{0},0) = (1 - \tanh x_{0})/6$ the breaking point is $\xi^{\ast}
 = 1$ (see figure~\ref{break}). 
\begin{figure}
\begin{center}
\includegraphics[width=6cm]{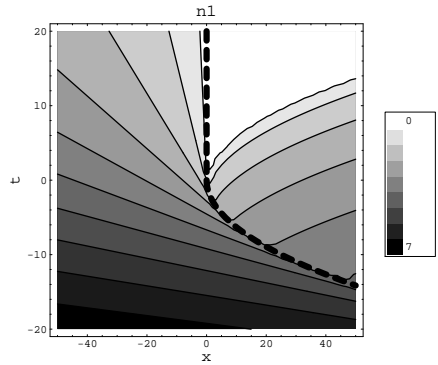}
\caption{\footnotesize{Density plot of the real part of the refractive
    index $n = n_{1} + i n_{2}$. Dashed line
    delimits the region where there is an imaginary contribution to
    refractive index (we denoted $t \equiv \xi$).}}
\label{realref}
\end{center}
\end{figure}
\begin{figure}
\begin{center}
\includegraphics[width=6cm]{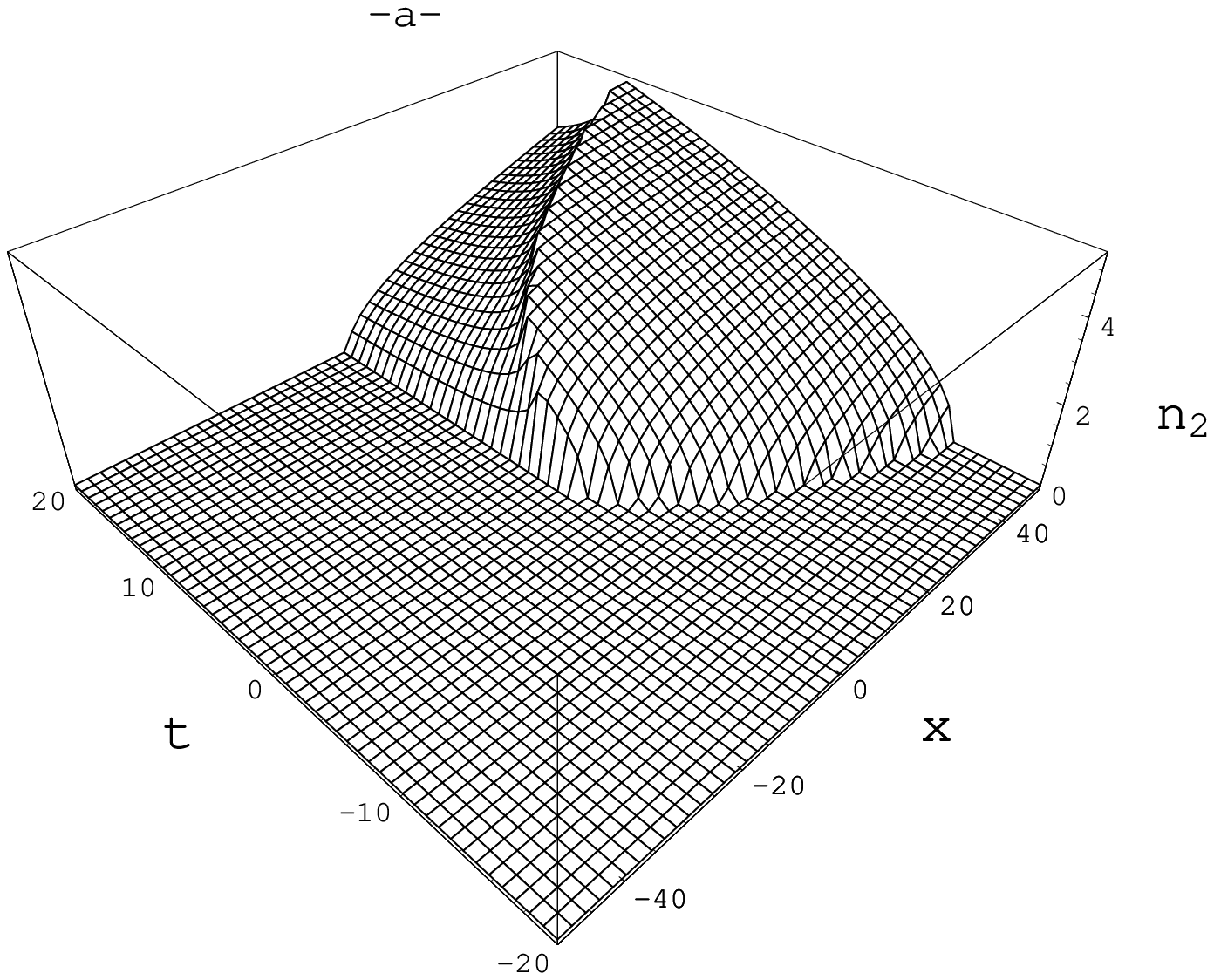}
\includegraphics[width=6cm]{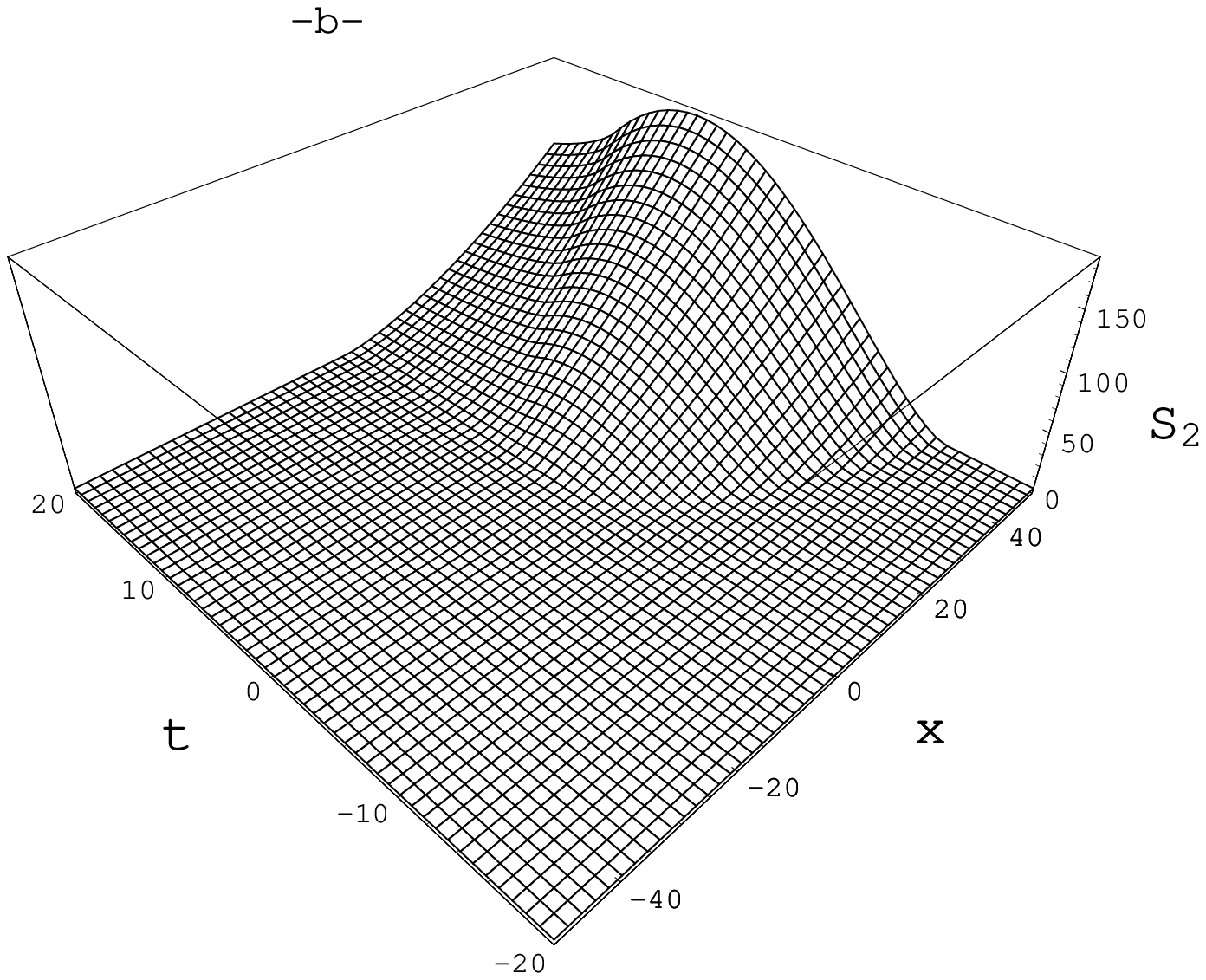}
\caption{\footnotesize{-a- Three-dimensional visualization of the
    imaginary part of
  refractive index. -b- Three-dimensional visualization of the
    imaginary part of
  $S$. It vanishes everywhere except inside the absorption region.}}
\label{surface}
\end{center}
\end{figure}

We remark that, as discussed in~\cite{kodkon}, the singular sector
of BH hierarchy induces a stratification of the affine space of
independent variables which gives rise to the integrable deformations
of hyperelliptic curves.

The propagation of the light on the plane $\pi : \xi = \xi_{0}$
($\xi_{0} = \text{const}$) is governed by the standard eikonal
equation~(\ref{dVN_phase1}). Denoting by $\Sigma$ the congruence
on $\pi$ normal to the family of curves $S(x,y, \xi_{0}) =
\lambda$, parametrized by $\lambda$, the curvature of $\gamma \in
\Sigma$ can be represented by the well known formula~\cite{Born}
\begin{equation}
\label{curvature1}
\kappa = \frac{1}{2 u} \; {\bf \nu} \cdot \nabla u,
\end{equation}
where $\nabla = \left(\partial_{x},\partial_{y} \right)$ and ${\bf
  \nu} = (\nu_{1}, \nu_{2})$ is the unit principal normal to $\gamma$
  on the plane $\pi$.
Now, let us observe that the curve $\gamma$ coincides locally with
the projection of the light
 ray on the plane $\pi$.
\begin{figure}
\begin{center}
\includegraphics[width=5cm]{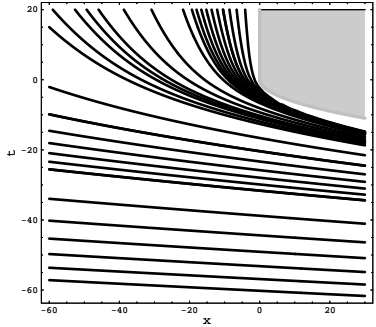}
\caption{\footnotesize{
    Wavefronts configuration on the plane $(x,\xi)$.}}
\label{wave}
\end{center}
\end{figure}

\noindent In virtue of the condition~(\ref{Dy}), the
formula~(\ref{curvature1}) assumes the form
\begin{equation}
\label{curvature2} \kappa = \frac{\nu_{1}}{2 u} \; u_{x}.
\end{equation}
Then, if $u_{x}$ blows up we conclude that the curvature of the
light ray blows up as well. This is the typical behavior which
light rays exhibit on the interface between
different materials. This type of solutions could be useful to
describe situations where light rays propagate in the
non-homogeneous medium and at certain point they cross some kind
of impurity. These impurities with drastically different optical
properties could be responsible of the abrupt change of direction.

As well known in electrodynamics, complex-valued refractive indices
also have a physical meaning since they can be used to describe
absorption effects of radiation. In
this specific case, complex values of the refractive index $n$ are
associated with a strong damping of the electromagnetic wave. We
illustrate this fact by the following explicit 
example.

Let us consider a solution obtained
from the hodograph relation~(\ref{hodograph}) setting
$\psi(u) = u^{2}$. Expliciting $u$ from~(\ref{hodograph}) and using
it in the expression~(\ref{phase}), one gets
\begin{align}
&u = \frac{1}{2} \left(- 6 \xi + \sqrt{36 \xi^{2} - 4 x} \right), \\
&S = \frac{4}{15} \sqrt{2} \sqrt{- 6 \xi + \sqrt{36 \xi^{2} - 4
x}} \left(6 \xi \left(- 6 \xi + \sqrt{36
\xi^{2}- 4 x} \right) + 12 x \right).
\end{align}
We note that the refractive index $n = 2 \sqrt{u}$ (we do not consider
negative refractive indices) is not real-valued on whole plane
$x$-$\xi$. Inside regions in which the refractive index takes a
complex contribution even the phase function $S$ becomes
complex-valued. Thus, writing $S = S_{1} + i S_{2}$, if $S_{2} > 0$ the electric field
acquires a damping factor
\begin{equation}
\label{damping}
{\bf E} = {\bf E}_{0} e^{- \omega S_{2}} \; e^{i \omega S_{1}}.
\end{equation}
In this regions strong absorptions effects are occurring.

Figure~\ref{realref} shows the real part of refractive
index $n = 2 \sqrt{u}$. The dashed line marks the region where the refractive
index acquires an imaginary contribution. This has been visualized
in the three-dimensional plot in the figure~\ref{surface}-a. Just where
refractive index
becomes complex-valued, the phase $S$ takes an imaginary
contribution indicating a strong
absorption (see figure~\ref{surface}-b).  \\
Figure~\ref{wave} shows the wave fronts configuration in
the plane $x$-$\xi$. Shaded region represents the absorption region.
For $\xi \to -\infty$ we have an almost plane wave front. Towards
absorption region, wave fronts deform itself following level lines
of refractive index. Just before absorption wave fronts present a
deviation of an almost right angle.

\end{document}